\documentclass[12pt,lathuile]{article}

\usepackage[dvips]{graphicx,color}
\usepackage{rotating}

\usepackage{epsfig}

\newcommand{\AmS}{{\protect\the\textfont2
  A\kern-.1667em\lower.5ex\hbox{M}\kern-.125emS}}
\newcommand{\pnn}{$K^+ \to \pi^+ \nu \bar\nu$}

\newcommand{\klpnn}{$K^0_L \to \pi^0 \nu \bar\nu$}
\newcommand{\klpnngen}{$K \to \pi \nu \bar\nu$}
\newcommand{\kpnn}{$K^+ \to \pi^+ \nu \bar\nu$}
\newcommand{\klpp}{$K_L \to \pi^0 \pi^0$}
\newcommand{\kpp}{$K^+ \to \pi^+ \pi^0$}

\newcommand{\kptwo}{$K^{+}\rightarrow\pi^{+}\pi^{0} $}

\newcommand{\kmn}{$K^{+}\rightarrow \mu^{+}\nu$}

\newcommand{\bklpnn}{$B(K^0_L \to \pi^0 \nu \bar\nu)$}

\newcommand{\bkpnn}{$B(K^+ \to \pi^+ \nu \bar\nu)$}
\newcommand{\kpnngen}{$K \to \pi \nu \bar\nu$}

\def\be{\begin{equation}}
\def\ee{\end{equation}}
\def\bea{\begin{eqnarray}}
\def\eea{\end{eqnarray}}

\def\vtd{V_{td}}

\begin{document}
\title{ 
MEASURING THE RARE DECAYS $K^+\to \pi^+ \nu \bar\nu$ and
$K^0_L \to \pi^0 \nu \bar\nu$ 
}
\author{
Milind V. Diwan         \\
{\em Brookhaven National Laboratory, Upton, NY 11973} 
}
\maketitle
\baselineskip=14.5pt
\begin{abstract}
In this paper I will describe the search for
rare kaon decay processes $K^+\to \pi^+ \nu \bar\nu$ and
$K^0_L \to \pi^0 \nu \bar\nu$. The
small decay rate for these  processes
is considered a key prediction of the standard model. 
We searched for the 
 charged kaon decay $K^+\to \pi^+ \nu \bar\nu$ 
using kaons decaying at rest in  the E787 detector at 
Brookhaven National Laboratory in two different phase 
space regions: Region 1 with pion momentum above 205 MeV/c
(the $K^+ \to \pi^+ \pi^0$ peak) and Region 2 
with pion momentum below 205 MeV/c. We have found 2 events in
Region 1 which is known to have small background
($0.15^{+0.048}_{-0.032}$).
This observation leads to a branching ratio measurement of 
 $(1.57^{+1.75}_{-0.82})\times 10^{-10}$.
We found 1 event in Region 2 with an expectation of 
$0.73\pm 0.18$ background events. The observation from Region 2 
is consistent with the branching ratio measured using Region 1. 
I will also describe new efforts underway to measure 
both the decay rates with much higher statistics.

\end{abstract}
\baselineskip=17pt
\newpage
\section{Introduction}

The rare kaon decays \pnn\ and \klpnn\ offer unique opportunities to
probe higher order phenomena associated with quark mixing and the
origin of CP non-invariance.  E787 at the Brookhaven
National Laboratory Alternating Gradient Synchrotron (AGS) has 
measured  the \pnn\ decay~\cite{pnn1} based on the observation of
two  events with low expected background. 
This measurement was based on data from the pion momentum region above 
205 MeV/c (the \kpp\ peak); we have recently extended this measurement
to the low momentum region with new techniques \cite{pnn2}.
 The branching ratio indicated by these
observations is consistent with the Standard Model (SM) expectation.
To fully explore the possibility of new physics or to make a precise
measurement of the t-d quark coupling $|\vtd |$ (assuming the SM level
for \bkpnn), a new measurement is about to commence. E949 is designed to
obtain a single event sensitivity of (8--14)$\times10^{-12}$, roughly
an order of magnitude below the SM prediction. In order to reach this
sensitivity the present detector has been upgraded, and
data running has begun with higher kaon flux.
With the completion of E949 in 2004, the possibility of an
inconsistency with the SM prediction of \pnn~ will be fully explored
or the important top-down quark mixing parameter will be determined to
a precision $15 - 30\%$ if the SM expectation is confirmed. 

\begin{figure}[t]
   \vspace{0cm}
$$
\epsfig{file=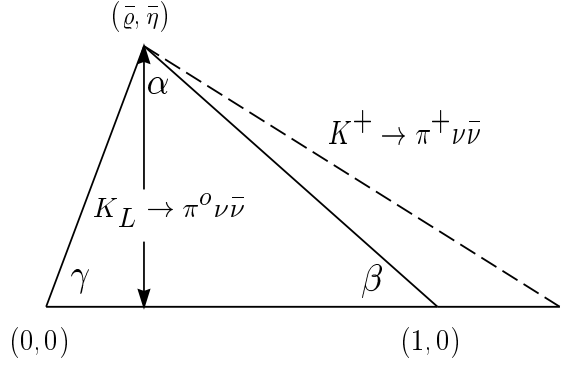,width=3in,height=2in,clip=t}
$$
   \vspace*{-0cm}
\caption{\label{triangle} The unitarity triangle.}
\end{figure}

	In addition, it has become
evident that the $K$ sector can yield the single most incisive
measurement in the study of CP violation through a measurement
of the branching ratio for \klpnn\ (\bklpnn), estimated to
be about $3 \times 10^{-11}$ . Within the SM this is a
unique quantity which directly measures the area of the CKM unitarity
triangles {\it{i.e.}} the physical parameter that characterizes all CP
violation phenomena, or the height of the triangle shown in
Fig.~\ref{triangle}.  The quest to observe \klpnn\ is being taken up
by the new KOPIO experiment at BNL discussed below.

The measurements of \bkpnn\ and \bklpnn\ will result in a complete picture of
Standard Model CP-violation in the $K$ system and a
comparison with comparably precise measurements anticipated from
the $B$ sector will be possible.

\section{Phenomenology of $K\to \pi \nu \bar\nu$. }

\begin{figure}[h]
   \vspace{0cm}
$$
\psfig{figure=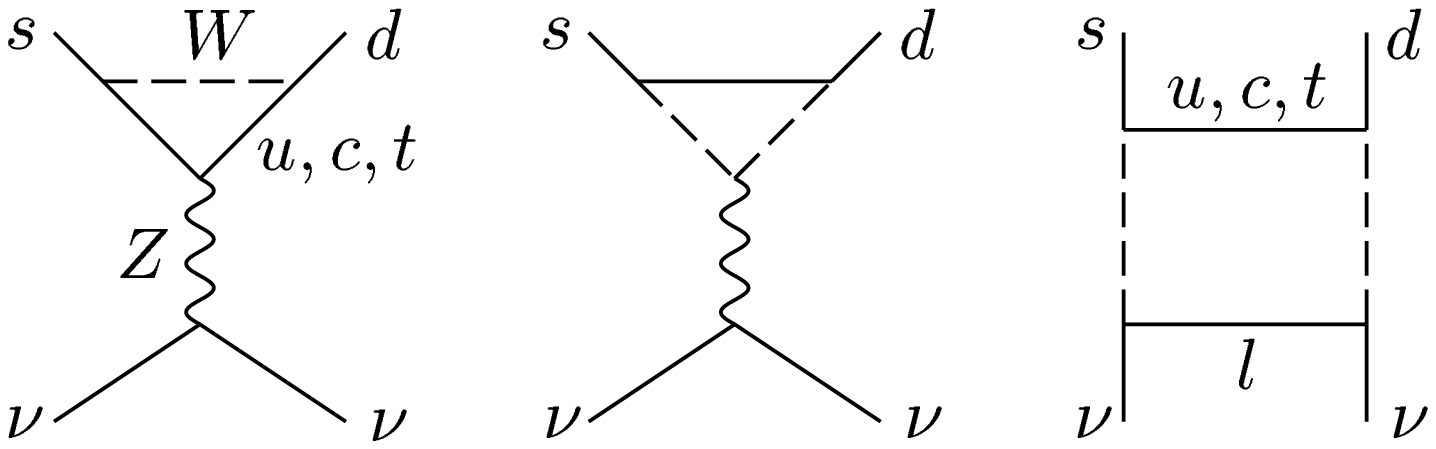,width=3in,
height=0.86in, clip=t}
$$
   \vspace*{-0cm}
\caption{\label{feyndiag} The leading electroweak diagrams
inducing \klpnngen~ decays. For \klpnn~ only the top quark contributes.}
\end{figure}

$K^+\to \pi^+ \nu \bar\nu$
 is a flavor-changing neutral current process, 
arising at the one loop level in the SM as shown in Fig.~\ref{feyndiag}.
The presence of the top quark in the loops makes this decay very
sensitive to the modulus of the elusive CKM coupling $\vtd$~\cite{smp}.
Moreover, this sensitivity can be fully exploited because of the hard
GIM suppression, the relatively small QCD corrections (which have been
calculated to next-to-leading-logarithmic order~\cite{Buchalla94}), and the 
fact that the normally problematic hadronic matrix element can be determined 
to a few percent from the rate of $K \to \pi e \nu$ ($K_{e3}$) 
decay~\cite{Marciano96}.  Taking account of all known contributions to the 
intrinsic theoretical uncertainty, the branching ratio can be calculated to 
a few percent~\cite{smp}, given the SM input parameters.  QCD corrections to
the charm contribution are the leading source of the residual
theoretical uncertainty.  Long distance contributions are known to be
negligible so not only can the effects of SM short-distance physics be
clearly discerned, but also the effects of possible non-SM physics.
In the SM, using current data on $m_t$, $m_c$, $V_{cb}$,
$|V_{ub}/V_{cb}|$, $\epsilon_K$, $\bar B - B$ mixing, etc., the
branching ratio is expected to be \bkpnn\ $= (0.75\pm 0.29)\times
10^{-10}$. 

The \klpnn~ decay mode is unique in that it is completely dominated by
direct CP violation~\cite{Littenberg89}
due to the CP properties of $K_L$, $\pi^0$ and
the relevant short-distance hadronic transition current. Since $K^0_L$
is predominantly a coherent, CP odd superposition of $K^0$ and $\bar
K^0$, only the imaginary part of $V^*_{ts} V_{td}$ survives in the
amplitude.  The comments made above about the hadronic matrix element,
QCD corrections, etc., in \kpnn\ also apply to \klpnn\ with the additional 
feature
that the lack of a significant charm quark contribution reduces the
intrinsic theoretical uncertainty to $\cal{O}$$(2\%)$.
Since the value of the sine of the Cabibbo angle is well
known ($|V_{us}| = \lambda = 0.2205$), $Im(V^*_{ts} V_{td})$ is equivalent
to the Jarlskog invariant, $\cal{J}$$ \equiv - Im(V_{ts}^* V_{td}
V_{us}^* V_{ud}) = -\lambda (1 - \frac{\lambda^2}{2}) Im(V_{ts}^*
V_{td})$.  $\cal{J}$, in turn, is equal to twice the area of any of
the six possible unitarity triangles~\cite{Jarlskog88}.  Since theoretical
uncertainties are extremely small, measurement of \bklpnn\ will
provide the standard against which all other measures of CP violation
will be compared, and even small deviations from the expectation
derived from SM predictions or from other measurements, {\it
e.g.} in the $B$ sector, will unambiguously signal the presence of new
physics. 
In the Wolfenstein parameterization of the CKM matrix,
\begin{equation}\label{bklpnw}
B(K^0_L\to\pi^0\nu\bar\nu)=1.8 \cdot 10^{-10} \eta^2 A^4 X^2(x_t)
\end{equation}
Inserting the current estimates for SM parameters into
Eq.~\ref{bklpnw}, the branching ratio for \klpnn~ is expected to be
in the range $(2.6 \pm 1.2)\cdot 10^{-11}$.  A clean measure of the
height of the unitary triangle, $\eta$, is provided by the \klpnn~
branching ratio.  We note that, all other parameters being known,
Eq. \ref{bklpnw} implies that the relative error of $\eta$ is half
that of \bklpnn.  Thus, for example, a $15\%$ measurement of \bklpnn~
can, in principle, determine $\eta$ to $7.5\%$.

 Most forms of new physics~\cite{bsmp, Leurer93, Grossman97} postulated to
augment or supersede the SM have implications for B(\kpnngen). 
In minimal supersymmetry and in some multi-Higgs
doublet models~\cite{Belanger92}, the extraction of $\sin 2\alpha$ and
$\sin 2\beta$ from CP asymmetries in B decays would be unaffected.
Such effects might then show up in a comparison with \klpnn, where,
e.g., charged Higgs contributions modify the top quark dependent
function in \bklpnn.  In other new physics scenarios,
such as supersymmetric flavor models~\cite{Nir98}, the effects in
\kpnngen~ tend to be small, while there can be large effects in the
$B$ (and also the $D$) system.  In these models the rare $K$ decays
are the only clean way to measure the true CKM parameters.
Loop diagrams involving new heavy particles in extensions of 
 the SM can interfere with SM diagrams
and alter the decay rate, and also  the 
kinematic  spectrum\cite{bsmp}.
 Exotic scenarios such 
$K^+ \to \pi^+ X^0$ where $X^0$ is a hypothetical 
stable weakly interacting particle or system of particles have also been 
suggested \cite{bs, aliev}.
 It is therefore 
important 
to obtain  higher statistics for the $K^+ \to \pi^+ \nu \bar\nu$ 
 decay and to extend the measurement 
to obtain the $\pi^+$ spectrum.


%

%
\section{E787 results}

\begin{figure}
\begin{center}
\epsfxsize 15cm
\epsffile{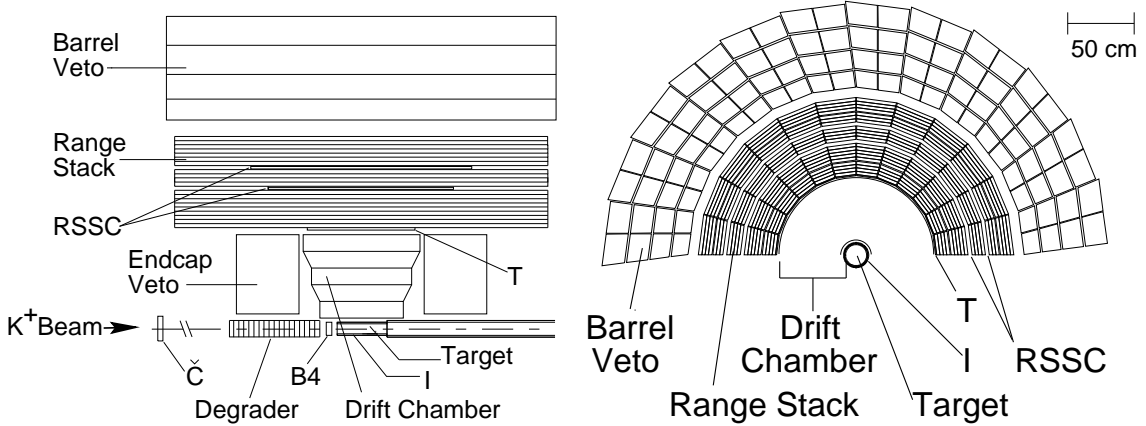}
\end{center}
\caption{Top half of side (left) and end (right) views of the E787 detector.}
\label{det}
\end{figure}

The signature for $K^+\to \pi^+ \nu \bar\nu$ 
in the E787 experiment (See Fig. \ref{det})
is a single  $K^+$ stopping in a   target (TG),
decaying 
 to a single $\pi^+$ with no other accompanying photons or charged particles.  
In Region 1, the major backgrounds were found  to be 
the two body decays  \kptwo\ ($K_{\pi2}$)
and  $K^+ \to \mu^+ \nu_\mu ~(K_{\mu2})$, 
scattered beam pions, 
and $K^+$ charge exchange (CEX) reactions resulting in 
decays $K^0_L \to \pi^+ l^- \bar\nu_l$, where $l = e$ or $\mu$. 
Region 2 has larger potential acceptance than Region 1 
because the  phase space is more than twice as large and the 
loss of pions due to nuclear interactions in the detector is smaller 
at the lower pion energies. However, there are additional sources of 
background for Region 2. These include $K_{\pi 2}$ in which the $\pi^+$ 
loses energy by scattering 
 in the material of the detector (primarily in the TG), $K^+ \to \pi^+ \pi^0 \gamma ~(K_{\pi2\gamma})$,
$K^+ \to \mu^+ \nu \gamma ~(K_{\mu2\gamma})$, 
$K^+ \to \mu^+ \nu \pi^0 ~(K_{\mu3})$,   
and
 $K^+ \to \pi^+ \pi^- e^+ \nu_e $ ($K_{e4}$) 
decays in which both the $\pi^-$ and the $e^+$ are 
invisible because of absorption. 

The data were obtained with a flux of 
$\sim 5\times 10^6$ electrostatically separated 
kaons per $\sim$1.6 sec spill at  700 to 800 MeV/c  
(with $\sim$20\% pion contamination)\cite{Dornboos00}
  entering the apparatus. The beam momentum and spill 
 was changed from time-to-time to get optimum efficiency. 
The kaons were identified  by a Cerenkov detector; 
two multi-wire-proportional-chambers were used to determine 
that there was only one entering particle. 
After slowing in a  BeO degrader  the kaons  traversed a 
10-cm-thick lead-glass (PBG) detector read out by 16 fine-mesh
photomultiplier tubes (PMT) and a scintillating  target hodoscope (TH)
placed before  the TG.
 The PBG detector  was designed to 
be insensitive to kaons and  
detect  electromagnetic showers  originating 
from kaon decays in the TG. 
The TH  was used to verify that there was only one kaon 
as well as determine 
the  position, time, and energy loss of the kaon before 
it entered and stopped in the TG. 
The TG  consisted of 413 5.0-mm-square, 
3.1-m-long plastic scintillating fibers, each connected to a PMT. 
The fibers were packed axially to form a cylinder of $\sim$12 cm  diameter.
Gaps in the outer edges of the TG were filled with smaller fibers 
which were connected to PMTs in groups. 
The PMTs were read out by ADCs, TDCs, and 
500 MHz transient digitizers based on GaAs charge-coupled 
devices (CCDs)\cite{ccd}. 
Photons  were detected in a hermetic calorimeter mainly 
 consisting of a 14-radiation length thick barrel detector made of 
lead/scintillator sandwich and 13.5-radiation length thick 
endcaps of undoped CsI crystals \cite{ec}.
The rest of the detector 
consisted of a central drift chamber (UTC)\cite{utc}, 
and a cylindrical 
range stack (RS)   of 21 layers of plastic 
scintillator with two layers of embedded tracking chambers, 
all within a 1-T solenoidal magnetic field. The TG, UTC, and RS 
allowed the measurement of the $P$, $R$, and $E$ 
of the charged decay products.  
The UTC had 12 layers of anode wires for measuring the transverse 
momentum and
six foils etched with helical cathode
strips to measure  dip angle or $z$  in the $r-z$ plane.
After  correction for energy loss in the target and
I-counter, the momentum resolution
was measured to be $\sigma_P/P\sim 1.1$\%. 
 After making corrections for the energy
loss in the sub-detectors before entering the
RS, the range and kinetic energy resolutions are measured to be 
$\sigma_R/R\sim 2.9$\% and $\sigma_E/\sqrt{E\mbox{(GeV)}}\sim 1.0$\%,
respectively.
The $\pi \to \mu \to  e$ 
decay sequence from pions that came to rest in the RS was observed using 
another set of 500 MHz transient digitizers (TD)\cite{td}.
The decay sequence observation is a powerful
tool in identifying a charged pion. Muon rejection from this information
can reach about $10^5$. This cut is
independent of the $\pi^+/\mu^+$ separation using another
 cut on the range
and momentum correlation for different particles, where the 
$\pi^+/\mu^+$ separation is more than 3$\sigma$.

The data reduction and offline  analysis for Region 1 and Region 2  were 
similar \cite{bergbush},
 although the final cuts 
to enhance signal and  suppress background to less than 
one event  were different.  
The UTC, RS with the TD readout and the photon veto system 
were the most important elements for Region 1 analysis. 
 The TG with the CCD pulse digitizing system, 
and the photon veto system
were  the important elements for Region 2 analysis. 
A multilevel trigger selected events 
by requiring 
 an identified $K^+$ to stop 
in the TG, followed,
after a delay of at least 1.5 ns, by a single charged particle track that
traversed TG and 
 RS  with a hit-pattern  
consistent with the expectation for $K^+ \to \pi^+ \nu \bar\nu$.
Events with photons were suppressed  by vetos on the barrel 
and endcap detectors. 
In the ~~offline analysis,
the single charged particle 
 was required to be identified as a $\pi^+$ with 
$P$, $R$, and $E$ consistent for a $\pi^+$, and 
the TD pulse information consistent with the decay sequence 
$\pi \to \mu \to e$  in  the 
last RS counter on the pion trajectory.
The kinematic variables  for Region 1 were corrected 
for various correlations for each event and the signal region
 was  defined   by the intervals
 $211 < P < 229 {\rm ~MeV/c}$,
$32.5 < R < 40.0 {\rm ~cm}$, and $115 < E < 135 {\rm ~MeV}$. 
For Region 2  the signal 
was defined by  $140 < P < 195 {\rm ~MeV/c}$,
$12 < R < 27 {\rm ~cm}$, and $60 < E < 95 {\rm ~MeV}$.

\subsection{Region 1 result} 

\begin{table}
\caption{Background estimates for 1995-97 and 1998 data.}
\label{bkg}
\vspace{0.4cm}
\begin{center}
\begin{tabular}[]{|l|l|l|}\hline
 Background   & 1995-7 & 1998\\ \hline
$\pi^+\pi^0$ & $0.0216 \pm 0.0050$& $0.0120 ^{+0.0031}_{-0.0042}$ \\ \hline
$\mu^+\nu_\mu$ &                    & $0.0092 \pm 0.0067$ \\
$\mu^+\nu_\mu\gamma$&                    & $0.0245 \pm 0.0155$ \\
$\mu^+\nu_\mu(\gamma)$
&$0.0282 \pm 0.0098$& $0.0337 ^{+0.0435}_{-0.0240}$ \\ \hline
1 beam bkg    & $0.0054 \pm 0.0042$& $0.0039 \pm 0.0012$ \\ \hline
2 beam bkg    & $0.0157 \pm 0.0149$& $0.0004 \pm 0.0001$ \\ \hline
CEX&$0.0096 \pm 0.0068$&$0.0157 ^{+0.0050}_{-0.0044}$ \\ \hline
Total  & $0.0804 \pm 0.0201$&$0.0657 ^{+0.0438}_{-0.0248}$ \\ \hline
\end{tabular}
\end{center}
\end{table}

The data sets for Region 1 were divided in two parts:
the 1995-97 data and the 1998 data. 
The 1995-97 data was already analysed and the results published (\cite{pnn1}),
nevertheless, we reanalysed the 1995-97 data with the same 
reconstruction programs as the 1998 data for consistency 
in the final results. We formed 
multiple  independent constraints  on each source of 
background. These constraints were grouped in two independent sets of cuts, 
designed to have  little  correlation.  One 
set of cuts was relaxed (or inverted) to enhance the background so that 
the other set could be evaluated to determine its power of rejection, as 
summarized below.
The  background due to  
\kmn\ was
obtained 
 by separately measuring the rejection factors 
 of the TD particle identification 
and kinematic ($R$ and $P$) particle identification. 
The background due to \kptwo\ was obtained by  
measuring the rejection factor of photon veto using the events 
in the kinematic peak of the two-body decay and the rejection factor 
of the kinematic cuts on events tagged by a photon.
The background 
from beam pions  was evaluated by 
separately measuring the rejections of Cerenkov,
  TH  beam particle identification, 
and the delay time between pion and kaon. 
The charge exchange background estimate is from Monte Carlo.
The kaon regeneration rate and beam profile are from the actual
measurement using data for the process of
$K^+n\rightarrow K^0_Sp, K^0_S\rightarrow\pi^+\pi^-$.
The final background estimate for Region 1 
 is shown in Table \ref{bkg}. 
The integrity of the background estimates was assured because 
the background cuts were defined  using only one-third of the data,
sampled uniformly from 
the entire set,
without examining the events in the pre-determined signal region. 
The cuts were then applied with no further changes
to the  remaining two-thirds of the data to obtain the final
background numbers.

\begin{figure}[t]
\begin{center}
\epsfxsize 8cm 
\epsffile{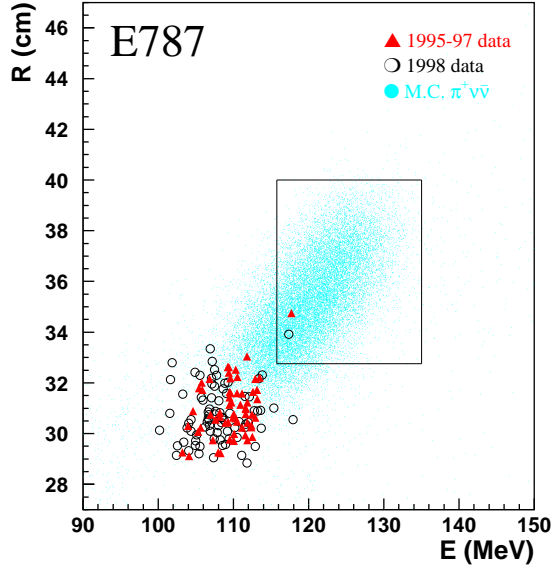}
\end{center}
\vspace{-0.3cm}
\caption{Range versus kinetic energy distribution for all E787
Region 1 
data with all cuts applied except for the range and kinetic energy
cuts.}
\label{box}
\end{figure}

Figure~\ref{box} shows the range versus kinetic energy for the events
surviving all the selection criteria. The box indicated by the solid
lines depicts the signal search region. Two signal events are found
inside this signal region and the events outside this box are
from the $K^+\rightarrow\pi^+\pi^0$ background due to photons escaping
 detection. Detailed studies of the candidate events  as well as a signal probability study showed that
they are consistent with the signature of
$K^+\rightarrow\pi^+\nu\bar{\nu}$.

The acceptance is estimated using $\mu^+\nu_\mu$, $\pi^+\pi^0$, and 
$\pi^+$-scattering monitor samples taken simultaneously with the
$\pi^+\nu\bar{\nu}$ trigger and by means of a 
Monte Carlo $\pi^+\nu\bar{\nu}$ sample.
Table~\ref{acc} gives the acceptances for each cut category and the
final acceptance. The corresponding branching ratio assuming 
no background is also in the table for the two different data sets.

\begin{table}
\caption{Acceptance study for $K^+\rightarrow\pi^+\nu\bar{\nu}$}
\label{acc}
\vspace{0.4cm}
\begin{center}
\begin{tabular}{|l|c|c|}\hline
Category                    & 1995-97 & 1998 \\ \hline\hline
$K^+$ stop efficiency&0.704     & 0.702 \\
$K^+$ decay after 2 ns&0.850 & 0.851 \\
$\pi^+\nu\bar\nu$ phase space&0.155     & 0.136 \\
Solid angle acceptance& 0.407     & 0.409 \\
$\pi^+$ nucl. interaction& 0.513     & 0.527 \\
Reconstruction efficiency& 0.959     & 0.969 \\
Other kinematic constraints& 0.665     & 0.554 \\
$\pi^+\rightarrow\mu^+\rightarrow e^+$ decay acc.
& 0.306     & 0.392 \\
Beam and target analysis& 0.699    & 0.706 \\
Accidental loss& 0.785     & 0.751 \\ \hline
Total acceptance&0.0021&0.0020 \\ \hline
Total $K^+$ triggers ($\times10^{12}$)&3.2 &2.7\\ \hline
Br($K^+ \rightarrow \pi^+ \nu \bar\nu$)
($\times 10^{-10}$) & $1.5^{+3.5}_{-1.2}$
                       & $1.9^{+4.4}_{-1.5}$\\ \hline
\end{tabular}
\end{center}
\end{table}

To combine searches with small statistics for both signal and
background level, a statistical analysis~\cite{junk} is performed
giving the final branching ratio measurement on
$K^+\rightarrow\pi^+\pi^0$ from E787: 
\begin{center}
Br($K^+ \rightarrow \pi^+ \nu\bar{\nu}$)=
                     $1.57^{+1.75}_{-0.82}\times 10^{-10}$
\end{center}
Assuming unitarity, $\bar{m}_t(m_t)=166\pm5$ GeV/$c^2$,
$M_W=80.41$ GeV/$c^2$ and $V_{cb}=0.041\pm0.002$, one can
derive the constraint
\begin{center}
$0.007<|V_{td}|<0.030$ (68\% C.L.),
\end{center}
without requiring any knowledge of $V_{ub}$ or $\epsilon_K$.

\subsection{Region 2 result}

For Region 2 we have so far analysed data from 1996 which
corresponded to a kaon exposure of $1.12\times 10^{12}$. 
The Region 2 background was found to be dominated by 
$K_{\pi 2}$ events in which the pion had a  
nuclear interaction near the kaon decay vertex, most probably 
on a carbon nucleus in 
the TG plastic scintillator. This scatter left
 the pion with  reduced 
kinetic energy, putting it in Region 2. 
We suppressed
 this background by removing events in which the pion track 
had a scattering signature in the TG. These signatures included 
kinks, tracks  that did not  point back to  the vertex fiber in which the 
kaon decayed,
 or energy deposits inconsistent with the 
ionization energy loss for a pion 
of the measured momentum. 
The remaining $K_{\pi2} $ background consisted of events in 
which the pion scattered  in one of the fibers 
 traversed by the kaon.  The extra energy deposits from the
pion scatters were obscured by the earlier large energy 
deposits of the kaon. For these events, 
we examined the  pulse shapes recorded in  the CCDs in each
 kaon  fiber  using a $\chi^2$ fit and eliminated  events in which   
an overlapping 
second pulse, in time with the pion, was found to have energy larger
 than 1 MeV. 
To obtain sufficient separation of the $K^+$  and $\pi^+$ induced pulses 
in the CCDs we  required a minimum
delay of 6 ns between the kaon and the pion. 
Finally, additional $K_{\pi2}$ rejection was obtained by removing 
events 
with photon interactions in detectors surrounding the kaon beam-line;
these cuts caused substantial ($\sim$ 42 \%) 
loss of efficiency because of accidental hits due to the high
flux of particles.

\begin{table} 
\begin{center}
\begin{tabular}{|l|l|r|}
\hline 
$K^+ \to \pi^+ \pi^0$ & d    &  $0.630\pm0.170$   \\
$K^+ \to \pi^+ \pi^0 \gamma$ & dm  &  $0.027\pm 0.004$  \\
$K_{\mu2\gamma} + K_{\mu3}$ & d &  $0.007\pm 0.007$  \\ 
Beam   & d &   $0.033\pm 0.033$   \\
$K^+ \to \pi^+ \pi^- e^+ \nu_e$ & dm &   $0.026\pm 0.032$       \\
CEX & dm &   $0.011\pm 0.011$    \\
\hline 
Total  &   &   $0.734\pm 0.177$    \\
\hline 
\end{tabular}
\end{center}
\caption{Estimated number of background events  
for  Region 2 of $K^+\to \pi^+ \nu \bar\nu$ data.
The second column indicates the method of background determination:
 data alone (d), data combined with simulation (dm).   
The errors include statistics of the data and  Monte Carlo as well as  
systematic uncertainties. \vspace{-0.5cm} }
\label{tab1}
\end{table}

\begin{figure}[t]
\begin{center}
\vspace{-2.0cm}
\epsfig{figure=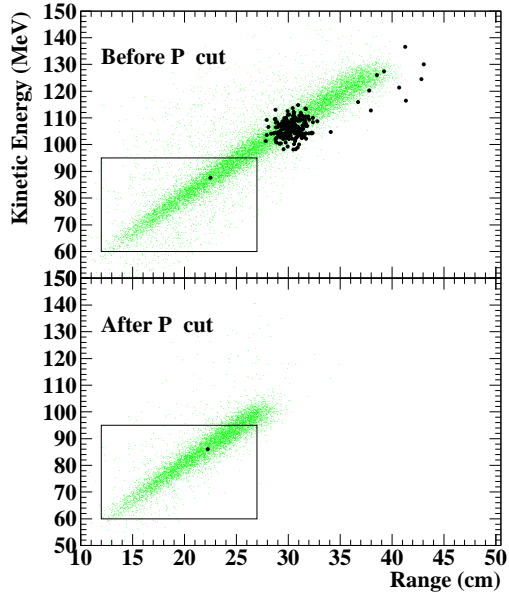, width=4in}
\vspace{-0.5cm} 
\end{center}
\caption{ 
Final plot for Region 2 data from 1996. 
Range (cm in plastic scintillator) and kinetic 
energy (MeV) of events remaining after all cuts 
except the momentum cut (top), and including the momentum cut (bottom).
 The dark points 
represent  the data. The simulated distribution of 
expected events from $K^+ \to \pi^+ \nu \bar\nu$
is indicated by the light dots.  The group of events around 108 MeV is due to the 
$K_{\pi2}$ background. The  events at higher energy are due
 to $K_{\mu2}$ and $K_{\mu2\gamma}$  background. 
All events except for the one in the signal box 
 are  eliminated by the $140 < P <195 {\rm ~MeV/c}$ 
cut on momentum. 
}
\label{rpeplt}
\end{figure}

The background estimates for Region 2 were performed using the 
same techniques as the Region 1 analysis: multiple independent constraints. 
The dominant background from $K_{\pi2}$ decay was measured by 
evaluating the rejection of the photon veto system on events tagged by
scattering signatures in the TG and target CCDs.
Similarly, the rejection of the target CCD cut was determined
 by using events 
that failed  the photon veto criteria.
It should be noted that  the Region 1 analysis measured 
 photon veto rejection using the unscattered events in 
the momentum peak (205 MeV/c). 
This method could not be used for Region 2 because the 
scattering in the TG
spoiled  the back-to-back correlation 
between the detected $\pi^+$ track in the RS and the undetected 
$\pi^0$, leading to different photon veto rejection factors  for scattered
and unscattered $K_{\pi2}$ background events.  
The other less important backgrounds were estimated using data or 
data combined with Monte Carlo estimates of cut efficiencies. 
 The final background estimates and associated errors in 
Table \ref{tab1} include   
corrections for    
small correlations  in the separate groups of cuts
and cross contamination of background samples. 
After all cuts the signal region for Region 2 was examined:
Fig.~\ref{rpeplt} shows the 
kinematics of the remaining events  
before and after the cut on measured momentum, $P$.

\begin{table}
\begin{center}
\begin{tabular}{|l|r|}
\hline 
Acceptance factors  &  \\
\hline 
$K^+$ stop efficiency & 0.670  \\
$K^+$ decay after 6 ns & 0.591 \\
$K^+ \to \pi^+ \nu \bar\nu$ phase space &  0.345  \\
Geometry   & 0.317  \\   
$\pi^+$ nucl. int. and decay in flight & 0.708  \\
Reconstruction efficiency &  0.957  \\    
Other kinematic cuts & 0.686   \\     
$\pi - \mu - e$ decay chain & 0.545  \\
Beam and target analysis  & 0.479  \\    
CCD acceptance  &  0.401   \\
Accidental loss &  0.363  \\
\hline 
Total acceptance & $7.65\times 10^{-4}$   \\
\hline 
\end{tabular}
\end{center}
\caption{Acceptance factors used in the measurement of 
$K^+\to \pi^+ \nu \bar\nu$  in Region 2. 
The ``$K^+$ stop efficiency'' is the fraction of kaons entering 
the TG that stopped.
``Other kinematic constraints'' include 
particle identification cuts.\vspace{-0.5cm}}
\label{accept}
\end{table}

\begin{figure}[t]
\begin{center}
\vspace{-0.5cm}
\epsfig{figure=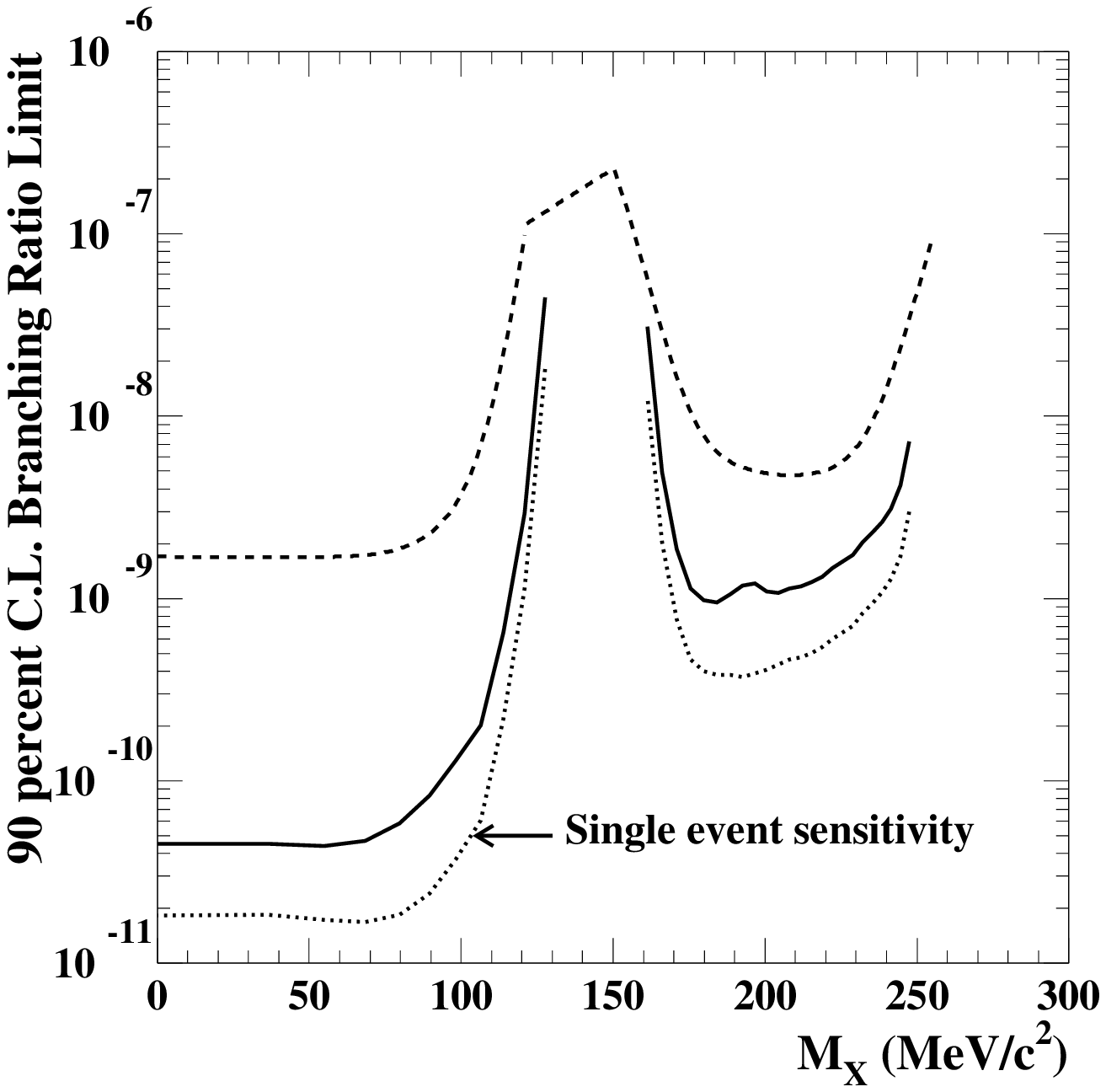, width=3in}
\vspace{-0.5cm}
\end{center}
\caption{ The 90\% C.L. upper limit for $B(K^+ \to \pi^+ X^0$) as a 
function of $M_{X^0}$, the mass of the recoiling system. 
The solid  (dashed) line is from this analysis (from \cite{pnn2early}).  
The limit for $M_{X^0}< 140 {\rm ~MeV/c^2}$ 
is derived from the result for Region 1. 
The observation of one event in Region 2 
causes a bump in the limit at $194 {\rm ~MeV/c^2}$.
Similarly the  2 events, consistent with the observation of 
$K^+ \to \pi^+ \nu \bar\nu$ above background,  
in Region 1 reported in \cite{pnn1}
increase  the limit at 
$105$ and $86 {\rm ~MeV/c^2}$.
We have also  
included  the single event sensitivity 
as a function of  $M_{X^0}$ (the dotted line), defined in the text, 
obtained by E787. }
\label{xplot}
\end{figure}

Using the total number of $K^+$ incident on TG for the 1996 data, 
 $1.12\times 10^{12}$, 
 the acceptance reported in Table \ref{accept}, and  the observation of one 
event in  Region 2  we calculate the
 upper limit of $B(K^+\to \pi^+ \nu \bar\nu) < 4.2\times 10^{-9}$ (90\% C.L.) 
\cite{feldman}. This is  
consistent with the branching ratio 
reported from Region 1 
and the SM decay spectrum \cite{pnn1}; combining the measurements from Region 1 and 
Region 2 does not alter the branching ratio measurement significantly because it is 
dominated by the sensitivity of Region 1. However,  
for non-standard scalar and tensor  interactions,
 Region 2 has  larger  acceptance  than Region 1. 
We have combined
the sensitivity of both regions to obtain 
the 90\% C.L. upper limits, $4.7\times 10^{-9}$ and 
$2.5\times 10^{-9}$, for scalar and tensor interactions, respectively. 

This measurement is also sensitive to $K^+ \to \pi^+ X^0$, 
where $X^0$ is a hypothetical 
stable weakly interacting particle, or system
 of particles. Fig.~\ref{xplot} shows 90\% C.L. upper 
limits on $B(K^+ \to \pi^+ X^0)$ together with
the previous limit from \cite{pnn2early}. 
The dotted line in Fig.~\ref{xplot} is 
the single event sensitivity defined as the
inverse of the acceptance for $K^+ \to \pi^+ X^0$ multiplied by the total 
number of stopped kaons as a
function of  $M_{X^0}$. 

The rest of the data from E787 is now under analysis for Region 2.
It is expected that the rest of the data should have lower 
background from \kptwo\ because of better performance from both 
the PBG detector and the CCD pulse digitizing system on the target.

\section{E949 and CKM}

E949 is based upon incremental upgrades to the techniques and
technology of E787. The extensive analysis of E787 data has been
used to project the sensitivity 
of E949. 
The net increase in sensitivity per year of E949 over initial E787
running is a factor of 13 coming from improvements summarized in
Table~\ref{tab_sens_factors}.
The most important detector upgrades are additional
 photon veto systems as well as improved trigger and data 
acquisition systems for better efficiency.
The largest factor in sensitivity, however, comes from the 
improved accelerator operations that will be optimized for E949.
The AGS will be operated
during the next several years in conjunction with RHIC
which requires injection roughly twice per
day leaving $\geq$ 20 hrs./day available for AGS slow extracted beam.
E949 will be the primary (or only) AGS user during the 2001-2003 period.
  After the proposed E949 running time of
6000 hours ($\sim$2 years or about 60 weeks), the expected \pnn\
sensitivity will be $1.7\times10^{-11}$.  Combined with the E787 data,
the result will reach $1.4\times10^{-11}$ with 0.7 expected background
events. 
As reported in this paper we now have much better understanding of
the background expected in Region 2. We have placed new photon 
veto systems in the upgraded E949 detector to reduce the 
background in Region 2 further.  
 With the added acceptance from  Region 2 below the \kpp, the
total 
sensitivity may reach $7.6\times10^{-12}$.  We would therefore expect
to see 7--13 events if the branching ratio is equal to the central SM
value of $10^{-10}$.

\begin{table}
\begin{center} \begin{tabular}{|l||c|} \hline Upgrade & Imprvmnt.\\
 & 
factor \\ \hline\hline Lower momentum & 1.44 \\ \hline Higher duty
factor & 1.53 \\ \hline E787 improvements & 1.54 \\ \hline Addtl.
efficiency improvements & 1.9 \\ \hline Rate effects \& below $K_{\pi2}$ & 2 \\
\hline\hline Total & 13 \\ \hline \end{tabular} \caption{Sensitivity/hr.
improvement factors for E949 compared to the 1995 run of E787.
}\label{tab_sens_factors} \end{center} \end{table}

\begin{figure}[t]
$$
\epsfig{figure=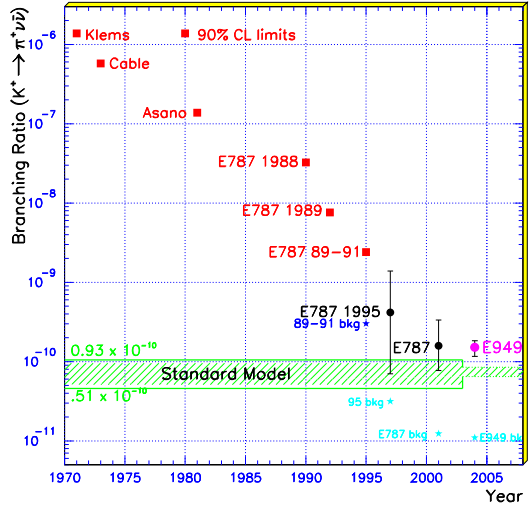,height=3in,width=3in}
$$
\caption{History of progress in the search for and measurement of
\kpnn. The solid squares represent single event sensitivities for
experiments that set limits on the branching ratio. The solid circles
represent the measured branching ratio for the E787
observation of this decay. The last  circle on right
 represents the expected
sensitivity for E949. The stars represent the achieved background level 
for E787  and the expected level for E949.}
\label{fig_history}
\end{figure}

Fig.~\ref{fig_history} is a plot of the progress in sensitivity for 
\kpnn\ experiments.
Included are experiments that set 90\% CL limits (earlier
E787 results and previous
experiments), as well as the single event
sensitivities for E787 and E949.

The E949 collaboration has physicists from BNL, FNAL,  UBC, TRIUMF, KEK, 
Alberta, Osaka, Fukui, IHEP-Moscow, INR-Moscow, Kyoto, UNM, RCNP.

An experiment (CKM) to measure $K^+ \to \pi^+ \nu \bar\nu$ using 
a very high flux of kaons in flight has recently been approved 
at Fermilab \cite{ckm}. 
The experiment relies on radio frequency separation techniques to 
obtain a clean 20 Gev/c kaon beam with intensity of $\sim$ 30 MHz. 
Many of the photon veto and background reduction techniques will be similar
to E787, however, CKM will employ a ring imaging Cerenkov 
detector to obtain additional measurement of the decay kinematics
to reduce background.   
The experiment intends to obtain $\sim$ 100 events 
with low background to make a precise measurement of this decay. 
An additional feature of CKM is that since it employs kaons decaying in 
flight, the main background in E787  for Region 2 should not 
present as much difficulty.
The detailed enumeration of backgrounds in Region 2 in E787 analysis 
is therefore important for CKM.

\section{KOPIO}

\begin{figure*}[htpb]
    \begin{minipage}{0.38\linewidth}
    \centerline{\epsfig{figure=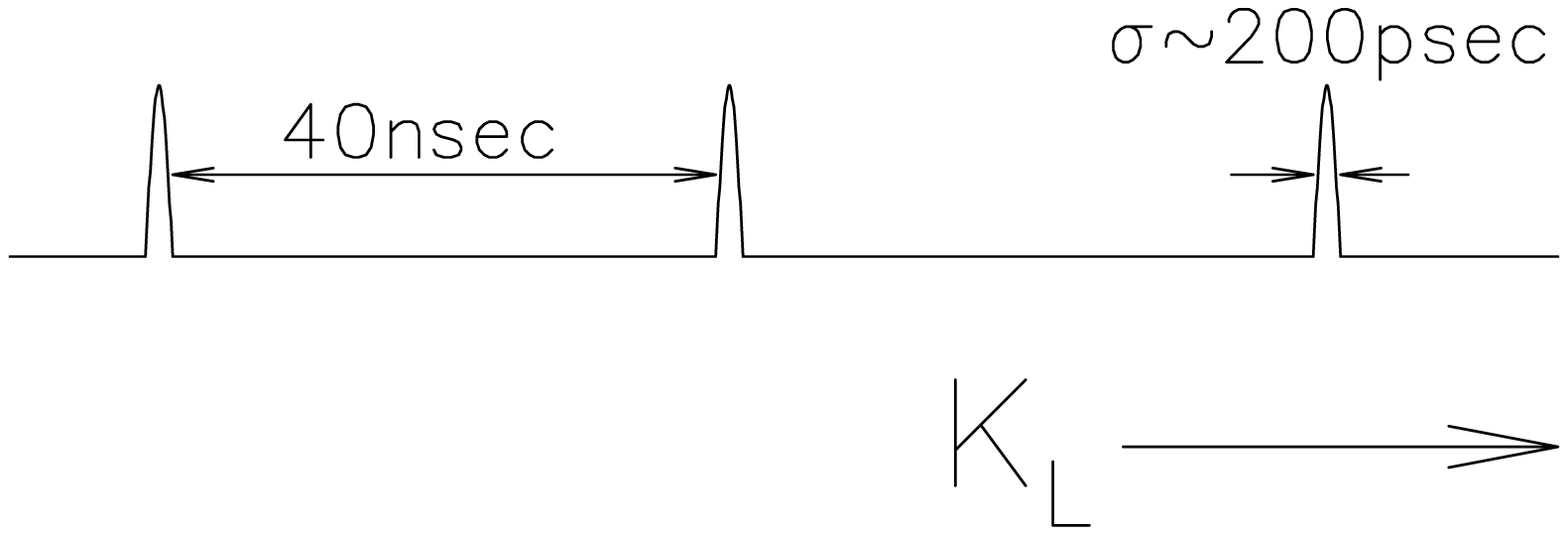,height=1.5in,width=2in}}
    \vspace{1in}
    \end{minipage}\hfill
\begin{minipage}{0.52\linewidth}
    \centerline{\epsfig{figure=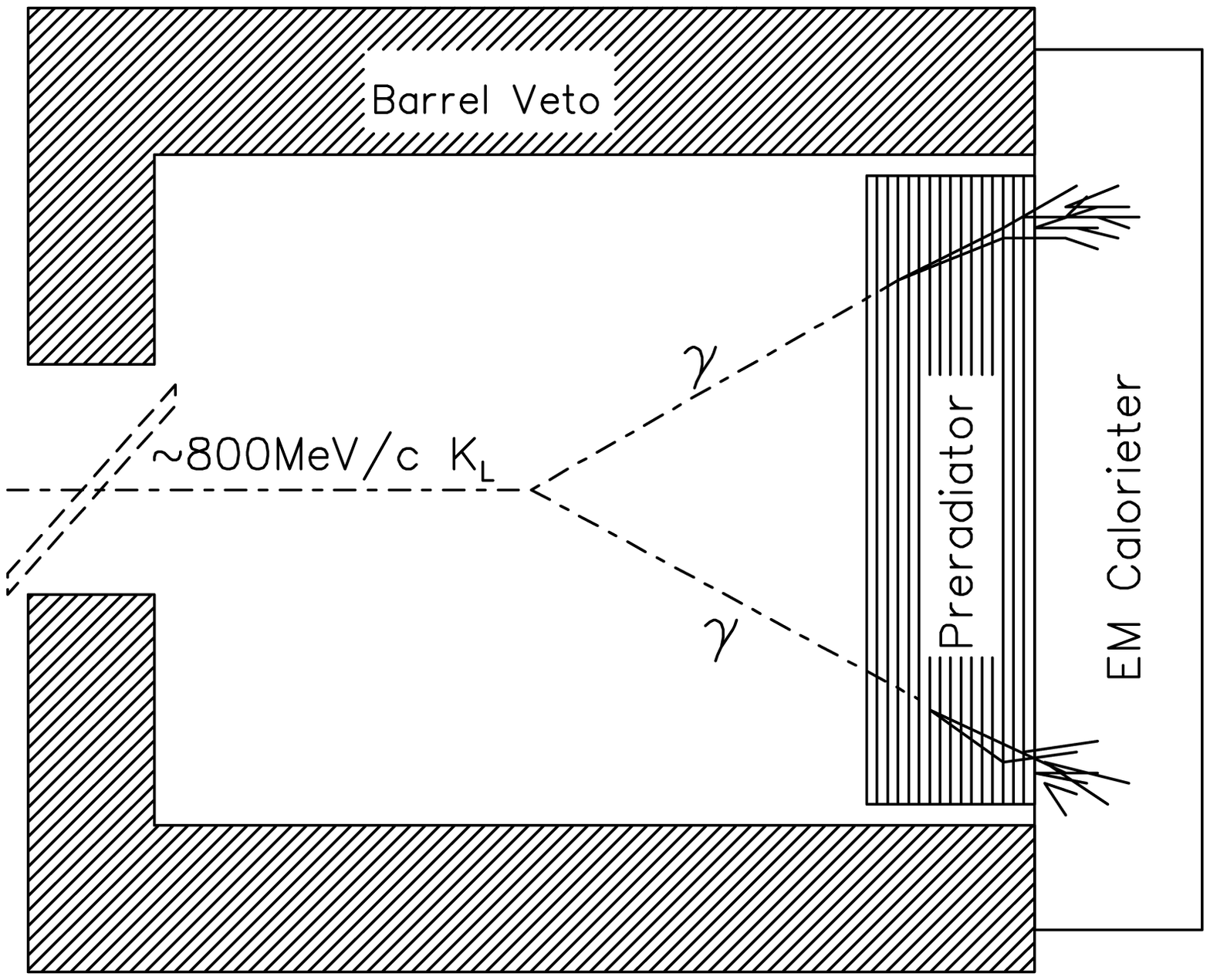,height=2in,width=3in}}
    \end{minipage}\hfill
    \caption{Elements of the KOPIO concept : a  pulsed primary beam
produces low energy kaons whose time-of-flight reveals their
momentum when the $\pi^0$ from \klpnn~ decay is reconstructed. }
    \label{f:schem}
\end{figure*}

Conclusive measurement of      \bklpnn\ , which is 
estimated to be $3\times 10^{-11}$,  
requires a
high kaon flux and very good background suppression.
The backgrounds mainly come from 
 competing decays 
that also yield $\pi^0$s  with branching ratios that are millions of
times larger, {\it e.g.} \klpp. 
Interactions between neutrons and kaons in the neutral beam
with residual gas in the decay volume can also result in emission of
single $\pi^0$s, as can the decays of hyperons which might occur in the
decay region, {\it e.g.} $\Lambda\to \pi^0 n$. 
The current experimental limit \bklpnn$< 5.9 \times
10^{-7}$~\cite{Alavi-Harati00} is a by-product of the KTeV experiment
at Fermilab.  It 
employed the Dalitz decay $\pi^0\to \gamma e^+ e^-$. 
 An experimental improvement
in sensitivity of more than four orders of magnitude is therefore required
to obtain the  signal for \klpnn~. 

The KOPIO experiment \cite{kopio} is designed to 
provide maximum possible redundancy for this kinematically
unconstrained 3 body decay; the experiment
 has an optimum system for ensuring that the
observed $\pi^0$ is the only detectable particle emanating from the
$K^0_L$ decay, and  has multiple handles for identifying possible
small backgrounds.
KOPIO employs a novel low energy, time-structured $K^0_L$ beam to allow
determination of the incident kaon momentum.  This intense beam, with
its special characteristics, can be provided only by the BNL
Alternating Gradient Synchrotron (AGS)~\cite{Glenn97}.
Utilizing low momentum also permits a detection system for the
$\pi^0$ decay photons that yields a fully constrained reconstruction
of the $\pi^0$'s mass, energy, and, momentum.  
 The photon veto system has features 
similar to those employed successfully in the E787 measurement of
\pnn  .

In 
Figure~\ref{f:schem} we show a simplified representation of the beam and
detector concept. The 24 GeV primary proton beam impinges on
the kaon production target in 200 ps wide pulses at a rate of 25 MHz
giving a microbunch separation of 40 ns. A 500 $\mu sr$ solid angle
neutral beam is extracted at $\sim 40^o$ to produce a ``soft'' $K_L$~
spectrum peaked at 0.65 GeV/c; kaons in the range from about 0.4 GeV/c
to 1.3 GeV/c are used.  The vertical acceptance of the beam (0.005 r)
is kept much smaller than the horizontal acceptance (0.100 r) so that
effective collimation can be obtained to severely limit beam halos and
to obtain an additional constraint on the decay vertex position.  Downstream
of the final beam collimator is a 4 m long decay region which is
surrounded by the main detector.  Approximately 16\% of the kaons
decay yielding a decay rate of about 14 MHz.  The beam region is
evacuated to a level of $10^{-7}$~Torr to suppress neutron-induced
$\pi^0$ production. The decay region is surrounded by an efficient
Pb/scintillator photon veto detector (``barrel veto'').  In order to
simplify triggering and offline analysis, only events with the
signature of a single kaon decay producing two photons occurring
within the period between microbunches are accepted.

Photons from \klpnn~ decay are observed in a two-stage endcap detector
comprised of a fine-grained preradiator followed by an 18 radiation
length (X$_0$) electromagnetic calorimeter.  The preradiator obtains
the times, positions and angles of the interacting photons from
$\pi^0$ decay by determining the initial trajectories of the first
$e^+ e^-$ pairs.  The preradiator consists of 64 0.034 X$_0$-thick
layers, each with plastic scintillator, converter and dual coordinate
drift chamber. The preradiator has a total effective thickness of 2
X$_0$ and functions to measure the photon positions and directions
accurately in order to allow reconstruction of the $K_L$ decay vertex
while also contributing to the achievement of sufficient energy
resolution.

The calorimeter located behind the preradiator consists of  ``Shashlyk''
tower modules, 11 cm by 11 cm in cross section and 18 X$_0$ in
depth. A Shashlyk calorimeter module consists of a stack of square
tiles with alternating layers of Pb and plastic scintillator read out
by penetrating WLS fibers. The preradiator-calorimeter combination is
expected to have an energy resolution of $\sigma_{\rm E}$/E$\simeq
0.033/\sqrt{{\rm E}}$.  Shashlyk is a proven technique which has been
used effectively in BNL experiment E865 and is presently the main
element in the PHENIX calorimeter at RHIC.

Suppression of most backgrounds is provided by a hermetic high
efficiency charged particle and photon detector system surrounding the
decay volume.  The system includes scintillators inside the vacuum
chamber, decay volume photon veto detectors and detectors downstream
of the main decay volume.  The barrel veto detectors are constructed
as Pb/scintillator sandwiches providing about 18 X$_0$ for photon
conversion and detection.  The detection efficiency for photons has
been extensively studied with a similar system in BNL experiment E787.
The downstream section of the veto system is needed to reject events
where photons or charged particles leave the decay volume through the
beam hole.  It consists of a sweeping magnet with a horizontal field,
scintillators to detect charged particles deflected out of the beam,
and photon veto modules.  A special group of counters - collectively,
the ``catcher'' - vetoes particles that leave the decay volume but
remain in the beam phase space.  This system takes advantage of the
low energy nature of our environment to provide the requisite veto
efficiency while being blind to the vast majority of neutrons and
$K^0$s in the beam. The catcher uses aerogel Cerenkov radiators read out with
phototubes.

The goal of KOPIO
is to obtain about 60 events with a signal to background ratio of
greater than 2:1.  This will yield a statistical uncertainty in the
measurement of the area of the CKM unitarity triangle of less than
10\%. While \klpnn~ is clearly the focus of KOPIO, many other
radiative $K$ decays of significant interest to the study of low
energy QCD will be measured and numerous searches for non-SM processes 
will also be conducted simultaneously.

The KOPIO collaboration includes UBC, BNL, Cincinnati, Kyoto, Moscow
(INR), New Mexico, TJNAF, TRIUMF, Virginia, VPI, Yale, and Zurich.
The experiment is scheduled to begin operation in 2005.

\section{Conclusion}

This work was supported by the U.S.  Department of Energy under
Contract No. DE-AC02-98CH10886.

\end{document}